\documentclass[%
 aip,
 jmp,%
 amsmath,amssymb,
 reprint,%
]{revtex4-1}

\usepackage{graphicx}% Include figure files
\usepackage{dcolumn}% Align table columns on decimal point
\usepackage{bm}% bold math
\begin{document}

\preprint{AIP/123-QED}

\title{Comparing the dark matter models, modified Newtonian dynamics and modified gravity in accounting for the galaxy rotation curves}% Force line breaks with \\

\author{Xin Li}
\email{lixin1981@cqu.edu.cn}
 \affiliation{Department of Physics, Chongqing University, Chongqing 401331, China}%Lines break automatically or can be forced with \\
 \affiliation{CAS Key Laboratory of Theoretical Physics, Institute of Theoretical Physics, Chinese Academy of Sciences, Beijing 100190, China}
\author{Li Tang}%
 \email{tang@cqu.edu.cn}
\affiliation{Department of Physics, Chongqing University, Chongqing 401331, China}%Lines break automatically or can be forced with \\

\author{Hai-Nan Lin}
 \email{linhn@ihep.ac.cn}
\affiliation{Department of Physics, Chongqing University, Chongqing 401331, China}%Lines break automatically or can be forced with \\
 %


\begin{abstract}
We compare six models (including the baryonic model, two dark matter models, two modified Newtonian dynamics models and one modified gravity model) in accounting for the galaxy rotation curves. For the dark matter models, we assume NFW profile and core-modified profile for the dark halo, respectively. For the modified Newtonian dynamics models, we discuss Milgrom's MOND theory with two different interpolation functions, i.e. the standard and the simple interpolation functions. As for the modified gravity, we focus on Moffat's MSTG theory. We fit these models to the observed rotation curves of 9 high-surface brightness and 9 low-surface brightness galaxies. We apply the Bayesian Information Criterion and the Akaike Information Criterion to test the goodness-of-fit of each model. It is found that non of the six models can well fit all the galaxy rotation curves. Two galaxies can be best fitted by the baryonic model without involving the nonluminous dark matter. MOND can fit the largest number of galaxies, and only one galaxy can be best fitted by MSTG model. Core-modified model can well fit about one half LSB galaxies but no HSB galaxy, while NFW model can fit only a small fraction of HSB galaxies but no LSB galaxy. This may imply that the oversimplified NFW and Core-modified profiles couldn't well mimic the postulated dark matter halo.
\end{abstract}

%Use showkeys class option if keyword
\maketitle

\section{Introduction}

It has long been found that rotation curves of spiral galaxies are significantly discrepant from the predictions of Newtonian theory \cite{Rubin:1978,Rubin1980,Bosma:1981,Sofue:2001}. According to Newton's law of gravitation, the gravitational force between two point-like particles is inversely proportional to the square of their separation. Therefore, the rotation velocity of a star far away from the galactic center is inversely proportional to the square root of the distance to the galactic center. However, the observations often show an asymptotically flat rotation curve out to the furthest data points \cite{Walter2008,Blok:2008}. There are several ways to reconcile this contradiction. The most direct assumption is that there is a large amount of nonluminous matter (dark matter) that has not been detected yet \cite{Begeman1991,Persic1996,Chemin:2011mf}. In fact, the dark matter hypotheses was first proposed to solve the mass missing problem of the sidereal system \cite{Kapteyn:1922,Oort:1932}. However, after decades of extensive research, no direct evidence for the existence of dark matter has been found on the particle physics level. This motivates us to search for other explanations of the discrepancy between the Newtonian dynamical mass and the luminous mass.

One possible way is to modify the Newtonian dynamics. In 1983 or so, M. Milgrom published a series of papers to modify the Newtonian dynamics in order to explain the flatness of galaxy rotation curves, which is well known today as the MOND theory \cite{Milgrom1983a,Milgrom1983b,Milgrom1986}. According to MOND, the Newton's second law no longer holds if the acceleration is small enough. The true dynamics should be $\mu(a/a_0){\bm a}={\bm a}_N$, where ${\bm a}_N$ is the acceleration in Newtonian theory, ${\bm a}$ is the true acceleration, and $a_0$ is the critical acceleration bellow which the Newtonian theory does not hold. The interpolation function $\mu(x)$ is chosen such that $\mu(x)\rightarrow 1$ when $a\gg a_0$, so the Newton's acceleration law is recovered. In the deep MOND region $a\ll a_0$, $\mu(x)\approx x$, such that the rotation curve keeps flat at large distance from the galactic center. MOND is a non-relativistic theory for a long time until the relativistic form is constructed by Bekenstein \cite{Bekenstein2004}. With only one universal parameter $a_0$, MOND has made great success in accounting for the rotation curves of spiral galaxies \cite{Sanders1996,Sanders:1998gr,Sander:2007,Swaters2010,Iocco:2015iia}.

In addition to modify the Newtonian dynamics, it is also possible to modify the Newtonian gravity (MOG). According to MOG, the Newton's law of gravitation is invalid at galactic scales. There are various MOG theories. Moffat \cite{Moffat:2005,Moffat:2006} proposed the scalar-tensor-vector gravity (STVG) and metric-skew-tensor gravity (MSTG) models, in which the gravitational ``constant" is no longer a constant, but is running with distance. Carmeli \cite{Carmeli1998,Carmeli2000,Carmeli2002} showed that the flatness of galaxy rotation curves can be naturally explained if the expansion of the universe is took into account, and argued that dark matter may be a intrinsic property of the spacetime. Horava \cite{Horava2009a,Horava2009b,Horava2009c} presented a candidate quantum field theory of gravity in ($3+1$) dimension spacetime, which is known as the Horava-Lifshitz theory. Grumiller \cite{Grumiller:2010bz} proposed an effective gravity whose potential contains a Rindler term in addition to the well known terms of general relativity. All of these theories can to a large degree reconcile the mass missing problem of galaxy rotation curves.

In this paper, we make a comprehensive comparison between different models in explaining the galaxy rotation curves. We choose 9 high-surface brightness (HSB) and 9 low-surface brightness (LSB) galaxies and fit the observed rotation curves to three different types of models, i.e. the dark matter, MOND and MOG models. To the dark matter models, we choose the NFW profile \cite{Navarro:1996,Navarro:1997} and the core-modified profile \cite{Brownstein:2009zz} for the dark matter halo. To the MOND models, we study Milgrom's MOND theory with two different interpolation functions. To the MOG models, we focus on the MSTG theory \cite{Moffat:2005}. We also compare these models to Newton's theory without adding the nonluminous dark matter. Thus there are six models in total. The best model to each galaxy is picked out using statistical method.

The outline of this paper is arranged as follows: In Section \ref{sec:models}, we introduce the theoretical models of galaxy structures and rotation velocities. In Section \ref{sec:results}, we introduce the data of 9 HSB and 9 LSB galaxies that are used in our fitting. We first obtain their surface brightness parameters by fitting to the photometric data, then obtain the model parameters by fitting to the observed rotation curve data. In Section \ref{sec:comparison}, we make the model comparison, and use the Bayesian Information Criterion (BIC) and the Akaike Information Criterion (AIC) to pick out the best model. Finally, discussion and summary are given in Section \ref{sec:summary}.

\section{theoretical models}\label{sec:models}

\subsection{Structure of galaxies}\label{sec:model_brightness}

The brightness of galaxy is often assumed to be a direct tracer of its mass distribution. The brightness of a HSB galaxy can in general be decomposed into two components, a ellipsoidal bulge and a flat disk. The bulge is usually modeled by an inhomogeneous ellipsoid with 3D spatial brightness \cite{Tamm:2005}
\begin{equation}
  l(a)=l_{0}\exp\left[-\left(\frac{a}{ka_{0}}\right)^{1/N}\right],
\end{equation}
where $l_{0}$ is the central density, $a_{0}$ is the harmonic mean radius of the bulge, $k$ is a normalization factor, $a=\sqrt{R^{2}+z^{2}/q^{2}}$ is the distance to the galactic center, $R$ and $z$ are the cylindrical coordinates, $q$ is the ratio of the minor axis to the major axis, and $N$ characterizes the shape of the profile. Integrating over $z$, we obtain the 2D surface brightness of the bulge,
\begin{equation}
  I_b(R)=2\int_R^{\infty}\frac{l(a)a}{\sqrt{a^2-R^2}}da.
\end{equation}
The thickness of the disk is very small compared to the galaxy size. We assume that the disk is infinitely thin, and model its surface brightness by the exponential law \cite{Vaucouleurs1959,Freeman1970},
\begin{equation}\label{eq:I_d}
I_{d}(R)=I_{0}\exp\left(-\frac{R}{h}\right),
\end{equation}
where $I_{0}$ is the central surface brightness in unit of $M_{\odot}~{\rm pc}^{-2}$, and $h$ is the scale length of the disk. We can equivalently convert Eq.~(\ref{eq:I_d}) to the logarithmic units using the relation
\begin{equation}
\mu~[{\rm mag}/{\rm arcsec}^{2}]=\mathcal{M_{\odot}}+21.572-2.5\log_{10} I~[L_{\odot}/{\rm pc}^{2}],
\end{equation}
and obtain
\begin{equation}
  \mu_d(R)=\mu_0+1.086\left(\frac{R}{h}\right),
\end{equation}
where $\mu_{0}$ is the central surface brightness in unit of ${\rm mag}~{\rm arcsec}^{-2}$, $\mathcal{M_{\odot}}$ and $L_{\odot}$ are the absolute magnitude and luminosity of the sun in a specific color-band. The total surface brightness of the HSB galaxies is given by $I(R)=I_b(R)+I_d(R)$. The free parameters are obtained by fitting $I(R)$ to the observed photometric data. We assume that the mass-to-light ratios of both the bulge and disk are constants.

For LSB galaxies, the surface brightness can be well approximated by the exponential disk as HSB galaxies, while the bulge component is usually negligible \cite{Blok:1995}. However, the gas in LSB galaxies is much richer than in HSB galaxies, so it is necessary to be considered. The mass profile of gas can be read from the observational data directly using the Groningen Image Processing System (GIPSY) whose home page is at http://www.astro.rug.nl/$\sim$gipsy/.

\subsection{Models of rotation velocity}\label{sec:model}

We first consider in the framework of Newtonian theory. The rotation velocity induced by the spheroidal bulge can be obtained by solving the Poisson equation, and lead to the result \cite{Tamm:2005}
\begin{equation}
V^{2}_{b}(R)=4\pi\sigma qG\int^{R}_{0}\frac{l(a)a^{2}}{\sqrt{R^{2}-e^{2}a^{2}}}da,
\end{equation}
where $G$ is Newton's gravitational constant, $\sigma$ is the mass-to-light ratio of the bulge, and $e=\sqrt{1-q^{2}}$ is the eccentricity of the bulge. Similarly, the rotation velocity induced by the infinitely thin exponential disk is given by \cite{Freeman1970}
\begin{eqnarray}
V^{2}_{d}(R) &=& \frac{GM}{2R}\left(\frac{R}{h}\right)^{3}\bigg{[}I_{0}\left(\frac{1}{2}\frac{R}{h}\right)K_{0}\left(\frac{1}{2}\frac{R}{h}\right) \nonumber\\[1mm]
&& -I_{1}\left(\frac{1}{2}\frac{R}{h}\right)K_{1}\left(\frac{1}{2}\frac{R}{h}\right)\bigg{]},
\end{eqnarray}
where $M=2\pi \tau h^{2}I_{0}$ is in total mass of the disk, $\tau$ is the mass-to-light ratio of the disk, $I_n$ and $K_n$ are the $n$th order modified Bessel functions of the first and second kinds, respectively. The rotation velocity due to the neutral hydrogen (HI) can be calculated from the mass profile of HI directly using GIPSY. We assume that the mass ratio of helium (He) to HI is $1/3$, and ignore other gases. Therefore, the rotation velocity contributes from the gas is given by
\begin{equation}
V^{2}_{\rm gas}=\frac{4}{3}V^{2}_{\rm HI}.
\end{equation}
Then rotation velocity arising from the combined contributions of bulge, disk and gas, can be written as the squared sum of each component, i.e,
\begin{equation}
V^{2}_{N}=V^{2}_{b}+V^{2}_{d}+V^{2}_{\rm gas},
\end{equation}
For the HSB galaxies, the gas component is negligible, hence $V_{\rm gas}=0$. For the LSB galaxies, the bulge component is negligible, hence $V_b=0$.

There are several models for the dark mater halo, such as the NFW profile \cite{Navarro:1996,Navarro:1997}, the pseudo-isothermal profile \cite{Jimenez:2003}, the Burkert profile \cite{Burkert:1995}, the Einasto profile \cite{Merritt:2006}, the core-modified profile \cite{Brownstein:2009zz}, and so on. All of these profiles can be generalized to the $(\alpha,\beta,\gamma)$-models \cite{Hemquist1990,Zhao:1996,An:2013}. Here we focus on the NFW profile and the core-modified profile. The density of NFW profile takes the form
\begin{equation}\label{eq:NFW}
\rho_{\rm NFW}=\frac{\rho_s r^{3}_{s}}{r(r+r_{s})^{2}},
\end{equation}
where $\rho_s$ and $r_{s}$ are the characteristic density and scale length, respectively. The mass of dark matter, which is acquired from the volumetric integration of Eq.~(\ref{eq:NFW}), contributes partly to the rotation curve,
\begin{equation}
V_{\rm s}^{2}=4\pi G\rho_{s} \frac{r^{3}_{s}}{r}\left[\ln\left(1+\frac{r}{r_{s}}\right)-\frac{r}{r+r_{s}}\right].
\end{equation}
Therefore, the rotation velocity in the NFW model is given by
\begin{equation}
V^{2}_{\rm NFW}=V^{2}_{N}+V_{\rm s}^{2}.
\end{equation}
NFW profile is often quantified by the viral radius $R_{\rm vir}$ and viral mass $M_{\rm vir}$ instead of $r_{s}$ and $\rho_{s}$ \cite{Navarro:1997,Wu:2008}. The viral radius $R_{\rm vir}$ is the radius within which the mean density of dark matter is 200 times the critical density $\rho_{\rm cr}$, and the viral mass $M_{\rm vir}$ is the mass of dark matter within $R_{\rm vir}$. These quantities are related by
\begin{equation}
\rho_{s}=\frac{200\rho_{\rm cr} R_{\rm vir}^{3}}{3r_{s}^{3}}\left[\ln\left(\frac{R_{\rm vir}+r_{s}}{r_{s}}\right)-\frac{R_{\rm vir}}{R_{\rm vir}+r_{s}}\right]^{-1},
\end{equation}
\begin{equation}
M_{\rm vir}=200\rho_{\rm cr}\frac{4}{3}\pi R_{\rm vir}^{3},
\end{equation}
where $\rho_{\rm cr}=3H_{0}^{2}/8\pi G$ is the critical density of the universe, and $H_{0}$=70 km s$^{-1}$ Mpc$^{-1}$ is the Hubble constant.

The NFW profile is singular at the galactic center. To avoid the singularity, Brownstein \cite{Brownstein:2009zz} proposed the so-called core-modified profile. The density of the core-modified profile takes the form
\begin{equation}\label{eq:core}
\rho_{\rm core}=\frac{\rho_{c}r^{3}_{c}}{r^{3}+r^{3}_{c}}.
\end{equation}
The mass of dark matter within the sphere of radius $r$ is given by
\begin{equation}\label{eq:core mass}
M(r)=\frac{4}{3}\pi \rho_{c}r^{3}_{c}\left[\ln\left(r^{3}+r^{3}_{c}\right)-\ln\left(r^{3}_{c}\right)\right].
\end{equation}
Thus, the corresponding rotation velocity is given by
\begin{equation}\label{eq:v}
V^{2}_{c}=\frac{4}{3}\pi G\rho_{c}\frac{r^{3}_{c}}{r}\left[\ln\left(r^{3}+r^{3}_{c}\right)-\ln\left(r^{3}_{c}\right)\right].
\end{equation}
Therefore, the rotation velocity in the core-modified profile is given by
\begin{equation}\label{eq:v core}
V^{2}_{\rm core}=V^{2}_{N}+V^{2}_{c}.
\end{equation}

According to MOND theory \cite{Milgrom1983a,Milgrom1983b}, the Newtonian dynamics is invalid when the acceleration is approaching or below the critical acceleration $a_0$. The effective acceleration is related to the Newtonian acceleration by
\begin{equation}\label{eq:g-gn-relation}
  \mu(g/a_0)g=g_N,
\end{equation}
where $g_N\equiv GM/r^2$ is the Newtonian acceleration, $a_0\approx 1.2\times 10^{-10}$ m s$^{-2}$ is the critical acceleration, and $\mu(x)$ is an interpolation function which has the asymptotic behaviors: $\mu(x)=x$ for $x\rightarrow 0$, and  $\mu(x)=1$ for $x\rightarrow \infty$. We choose two interpolation functions, the first one is the standard interpolation function initially proposed by Milgrom \cite{Milgrom1983a}
\begin{equation}\label{eq:interp-function}
  \mu_{1}(x)=\frac{x}{\sqrt{1+x^2}}.
\end{equation}
Combining Eq.~(\ref{eq:g-gn-relation}) and Eq.~(\ref{eq:interp-function}), we can solve for $g$,
\begin{equation}
  g^2=\frac{1}{2}g_N^2\left(1+\sqrt{1+\left(\frac{2a_0}{g_N}\right)^2}\right).
\end{equation}
Since $V=\sqrt{gR}$ and $V_N=\sqrt{g_NR}$, we obtain the rotation velocity in the MOND theory
\begin{equation}
V^{2}_{\rm MOND1}=\sqrt{\frac{V^{4}_{N}}{2}+\sqrt{\frac{V^{8}_{N}}{4}+R^{2}a^{2}_{0}V^{4}_{N}}}.
\end{equation}
Another widely used interpolation function is the so-called simple interpolation function\cite{Famaey:2005}
\begin{equation}\label{eq:interp-function2}
  \mu_{2}(x)=\frac{x}{1+x}.
\end{equation}
This interpolation function can give a better fit to the Milky-Way-like HSB galaxies than the standard interpolation function \cite{Famaey:2005,Zhao:2006}.
The corresponding rotation velocity is given by
\begin{equation}
V^{2}_{\rm MOND2}=\frac{V^{2}_{N}+\sqrt{V^{4}_{N}+4RaV^{2}_{N}}}{2}.
\end{equation}

The MSTG model is presented by Moffat \cite{Moffat:2005}. The action of MSTG model is the Einstein-Hilbert action $S_{EH}$ added by a mass term $S_M$, a scale field term $S_F$, and a term characterizes the interaction between mass and scale field $S_{FM}$. In the linear weak field approximation, the MSTG acceleration law of test particles reads
\begin{equation}\label{eq:a_MSTG}
a(R)=-\frac{G_{N}M}{R^{2}}\left\{1+\sqrt{\frac{M_0}{M}}\left[1-\exp(-R/r_{0})\left(1+\frac{R}{r_{0}}\right)\right]\right\},
\end{equation}
where $G_{N}$ is the Newton's gravitational constant, $M$ is the mass of the particle, $M_0$ and $r_{0}$ are characteristic parameters. The best fitting to a large amount of galaxy rotation curves shown that both $M_0$ and $r_{0}$ are approximately universal constants, i.e. $M_0\approx 9.6\times10^{11}M_{\odot}$ and $r_0\approx 13.92$ kpc \cite{Brownstein:2006zz}. Eq.~(\ref{eq:a_MSTG}) can be regarded as the Newtonian acceleration except that the Newton's gravitational constant is replaced by the running gravitational ``constant"
\begin{equation}
  G(R)=G_N\left\{1+\sqrt{\frac{M_0}{M}}\left[1-\exp(-R/r_{0})\left(1+\frac{R}{r_{0}}\right)\right]\right\}.
\end{equation}
Therefore, the rotation velocity in the MSTG model is given by
\begin{equation}
V^{2}_{\rm MSTG}=V^{2}_{N}G(R)/G_N.
\end{equation}

\section{Best-fitting results}\label{sec:results}

\subsection{Best-fitting to the surface brightness}\label{sec:brightness}

Our samples consists of 9 HSB galaxies and 9 LSB galaxies taken from published literatures. All the 9 HSB galaxies are taken from Ref.\cite{Palunas:2000}, and the surface brightness are imaged at I-band. As for the LSB galaxies, 8 of them are imaged at R-band, and the rest one (F730-V1) are imaged at V-band.
Our samples are the same to that in Kun et al.\cite{Kun:2016yys}. We fit the photometric data to the models discussed in Section \ref{sec:model_brightness} using the least-$\chi^2$ method. HSB galaxies are fitted by a bulge plus a disk, while LSB galaxies are fitted by a disk only. We list the surface brightness parameters in Table~\ref{tab:HSB_brightness} and Table~\ref{tab:LSB_brightness}for HSB galaxies and LSB galaxies, respectively.
\begin{table*}[htbp]
%\begin{ruledtabular}
\centering                   %表格居中              % 默认字体大小
\caption{\small{The surface brightness parameters of HSB galaxies. The photometric data are taken from Ref.\cite{Palunas:2000}.}} %“table” 加粗，特殊字符、希腊字母需要加$ $          %对应前面，\\换行
%\vskip 0.1in                %表示上下间隔
%\arrayrulewidth=1.0pt          %线宽
\renewcommand{\arraystretch}{1.3}   %行宽
\resizebox{!}{3.2cm}
{\begin{tabular}{ccccccc} % creating  columns
\hline\hline % inserting double-line
& \multicolumn{4}{c}{Bulge}  & \multicolumn{2}{c}{Disk}\\
\cline{2-7}
            &$l_{0}$            & $ka_{0}$          & $N$               & $q$               & $\mu_{0}$ & $h$ \\
            &$L_{\odot}/{\rm pc}^{3}$ & kpc         &                   &                   & ${\rm mag}/\rm {arcsec}^{2}$ & kpc \\ \hline
ESO215G39   &0.7805$\pm$0.0080	&0.6000$\pm$0.0027	&0.6684$\pm$0.0042	&0.7805$\pm$0.0080	&19.3033	&3.4302\\
ESO322G76	&1.4166$\pm$0.0089	&0.6272$\pm$0.0017	&0.7397$\pm$0.0022	&0.7082$\pm$0.0044	&18.9113	&2.8937\\
ESO322G77	&5.7094$\pm$0.0103	&0.1920$\pm$0.0008	&1.0054$\pm$0.0030	&0.5709$\pm$0.0010	&18.7573	&2.0541\\
ESO323G25	&1.8523$\pm$0.0270	&0.4706$\pm$0.0015	&0.0212$\pm$0.0039	&0.3346$\pm$0.0053	&18.5259	&3.0958\\
ESO383G02	&6.3404$\pm$0.0162	&0.3573$\pm$0.0004	&0.7261$\pm$0.0009	&0.9512$\pm$0.0024	&19.7883	&4.9680\\
ESO445G19	&2.1046$\pm$0.0186	&0.3723$\pm$0.0014	&0.7404$\pm$0.0034	&0.7368$\pm$0.0065	&19.3500	&4.3200\\
ESO446G01	&1.0082$\pm$0.0019	&0.6902$\pm$0.0031	&1.1141$\pm$0.0030	&0.8982$\pm$0.0016	&19.7785	&5.6008\\
ESO509G80	&2.8310$\pm$0.0209	&0.7220$\pm$0.0023	&0.7049$\pm$0.0028	&0.2831$\pm$0.0021	&19.2745	&5.4382\\
ESO569G17	&2.8078$\pm$0.0038	&0.3645$\pm$0.0011	&0.7965$\pm$0.0025	&0.8985$\pm$0.0012	&18.2900	&1.6900\\ \hline
\hline% inserts single-line
\end{tabular}}\label{tab:HSB_brightness}
\end{table*}

\begin{table}[htbp]
\centering                  %表格居中
\caption{\small{The surface brightness parameters of LSB galaxies. The references are given in the last column.}} %“table” 加粗，特殊字符、希腊字母需要加$ $
%\vskip 0.1in                %表示上下间隔
%\arrayrulewidth=1.0pt          %线宽
\renewcommand{\arraystretch}{1.3}   %行宽
\resizebox{!}{3.2cm}
{\begin{tabular}{lccc} % creating  columns
\hline\hline % inserting double-line
& \multicolumn{3}{c}{Disk}\\
\cline{2-4}
& $\mu_{0}$ & $h$ &Reference \\
& ${\rm mag}/{\rm arcsec}^{2}$ &kpc & \\ \hline
F561-1	&22.39	&2.60 &Ref.\cite{Hulst:1993}\\
F563-1	&22.47	&2.60 &Ref.\cite{Blok:1996}\\
F568-3	&22.17	&3.40 &Ref.\cite{Blok:1995}\\
F579-V1	&21.90	&3.80 &Ref.\cite{Blok:1996}\\
F583-1	&22.01	&1.10 &Ref.\cite{Blok:1996}\\
F730-V1	&21.75	&6.00 &Ref.\cite{Kim:2007}\\
UGC128	&22.38	&5.10 &Ref.\cite{Blok:1995}\\
UGC1230	&22.54	&3.40 &Ref.\cite{Hulst:1993}\\
UGC5750	&21.80	&4.20 &Ref.\cite{Blok:1995}\\ \hline
\hline% inserts single-line
\end{tabular}}\label{tab:LSB_brightness}
\end{table}

\subsection{Best-fitting to the galaxy rotation curves}\label{sec:best-fitting}

We fit the observed rotation curve data to the theoretical models discussed in Section \ref{sec:model} using the least-$\chi^2$ method. The best-fitting parameters are obtained by minimizing the $\chi^2$,
\begin{equation}
\chi^{2}=\sum^{n}_{i=1}\left[\frac{V_{\rm th}(r_{i})-V_{\rm obs}(r_{i})}{\sigma_{i}}\right]^{2},
\end{equation}
where V$_{\rm th}$ is the theoretical velocity, V$_{\rm obs}$ is observed velocity, and $\sigma$ is the $1\sigma$ error of V$_{\rm obs}$. In the Baryonic model and MOND models, the only two free parameters are the mass-to-light ratios of the bulge ($\sigma$) and disk ($\tau$). For LSB galaxies, $\sigma\equiv 0$, and there is only one free parameter. The critical acceleration in the MOND models is fixed at $a_{0}=1.2\times10^{-13}~{\rm km~s}^{-2}$ \cite{Begeman1991}. In the dark matter models, there are two additional parameters, i.e. $M_{\rm vir}$ and $R_{\rm vir}$ in the NFW model, and $\rho_c$ and $r_{c}$ in the core-modified model. In the MSTG models, there are also two additional parameters, i.e. the characteristic mass $M_0$ and scale length $r_{0}$. However, we find that theses two parameters couldn't be well constrained using our galaxy sample. Therefore, we fix them to the values $M_0=9.6\times10^{11}M_{\odot}$ and $r_0=13.92$ kpc, which are obtained from fitting to a large sample of galaxies and taking the average \cite{Brownstein:2006zz}.

\begin{figure*}[htbp]
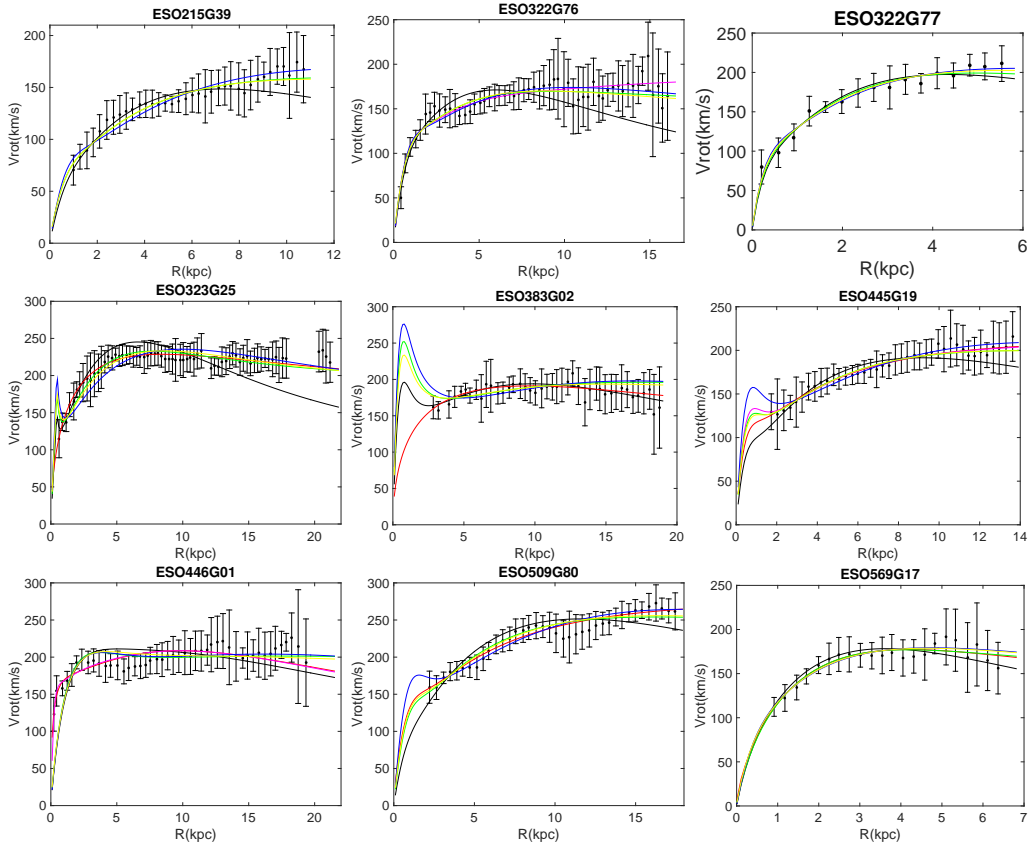


\includegraphics[width=0.32\textwidth]{ESO215G39.eps}
\includegraphics[width=0.32\textwidth]{ESO322G76.eps}
\includegraphics[width=0.32\textwidth]{ESO322G77.eps}
\includegraphics[width=0.32\textwidth]{ESO323G25.eps}
\includegraphics[width=0.32\textwidth]{ESO383G02.eps}
\includegraphics[width=0.32\textwidth]{ESO445G19.eps}
\includegraphics[width=0.32\textwidth]{ESO446G01.eps}
\includegraphics[width=0.32\textwidth]{ESO509G80.eps}
\includegraphics[width=0.32\textwidth]{ESO569G17.eps}
\caption{\label{fig:HSB} The best-fitting rotation curves of HSB galaxies in different models: black solid curves for Baryonic model, red solid curves for NFW model, magenta solid curves for core-modified profile model, green solid curves for standard MOND model, yellow solid curves for simple MOND model, and blue solid curves for MSTG model. The black dots with $1\sigma$ error bars are the observed data.}
\end{figure*}

The rotation curve data of HSB galaxies are taken from Palunas \cite{Palunas:2000}. We list the best-fitting parameters in Table~\ref{tab:HSB_parameter}. We also list the reduced chi-square $\chi^{2}/{\rm dof}$, where ${\rm dof}=N-p$ is the degree of freedom, $N$ is the number of data points and $p$ is the number of free parameters. For three galaxies (ESO215G39, ESO322G76 and ESO322G77) in NFW model and five galaxies(ESO215G39, ESO322G77, ESO323G25, ESO509G80, ESO569G17) in the core-modified model, the best-fitting scale parameters ($r_s$ and $r_c$) of the dark matter halo overstep the galaxy scale, which is physically unreasonable. The mass-to-light ratio of disk for ESO383G02 in the core-modified model is unphysically small. Therefore, we do not list them in Table~\ref{tab:HSB_parameter}. For four HSB galaxies (ESO323G25, ESO383G02, ESO446G01, ESO569G17), the mass-to-light ratio of the bulge couldn't be well constrained in the NFW model, and we fix it to be zero. The best-fitting curves accompanied by the observed data are plotted in Fig.~\ref{fig:HSB}. The error bar represent the $1\sigma$ uncertainty.

\begin{figure*}[htbp]
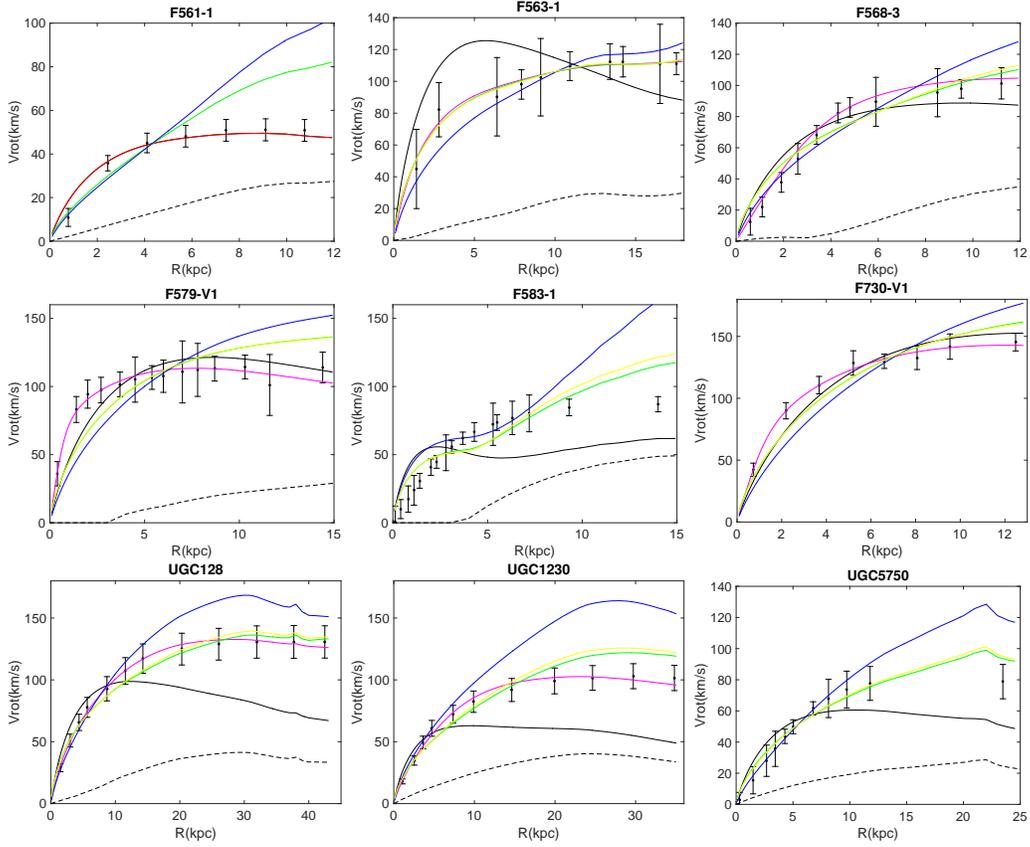


\includegraphics[width=0.32\textwidth]{F561-1.eps}
\includegraphics[width=0.32\textwidth]{F563-1.eps}
\includegraphics[width=0.32\textwidth]{F568-3.eps}
\includegraphics[width=0.32\textwidth]{F579-V1.eps}
\includegraphics[width=0.32\textwidth]{F583-1.eps}
\includegraphics[width=0.32\textwidth]{F730-V1.eps}
\includegraphics[width=0.32\textwidth]{UGC128.eps}
\includegraphics[width=0.32\textwidth]{UGC1230.eps}
\includegraphics[width=0.32\textwidth]{UGC5750.eps}
\caption{\small{The best-fitting rotation curves of LSB galaxies in different models: black solid curves for Baryonic model, magenta solid curves for core-modified profile model, green solid curves for standard MOND model, yellow solid curves for simple MOND model, and blue solid curves for MSTG model. The black dots with $1\sigma$ error bars are the observed data. The black dashed curves are the contributions from the gas. Curves for NFW model are not shown here.}}\label{fig:LSB}
\end{figure*}

The rotation curve data of LSB galaxies are taken from different literatures. F579-V1 is taken from Blok \cite{Blok:1996}, UGC128 and UGC1230 are taken from Hulst \cite{Hulst:1993}, and the rest five galaxies are taken from McGaugh \cite{McGaugh:2001}. We list the best-fitting parameters in Table~\ref{tab:LSB_parameter}. For all the LSB galaxies in NFW model and three LSB galaxies (F561-1, F583-1, UGC5750) in core-modified model, the parameters couldn't be well constrained. The mass-to-light ratio of disk for F568-3 in the core-modified model is unphysically small. Therefore we do not list them here. The best-fitting curves accompanied by the observed data are plotted in Fig.~\ref{fig:LSB}, where the contributions from the gas are also shown with black dashed curves.

\begin{table*}[htbp]
\centering                  %表格居中
\caption{\small{The best-fitting parameters of HSB galaxies in different models. For three galaxies (ESO215G39, ESO322G76 and ESO322G77) in NFW model and five galaxies (ESO215G39, ESO322G77, ESO323G25, ESO509G80, ESO569G17) in the core-modified model, the best-fitting scale parameters ($r_s$ and $r_c$) of the dark matter halo overstep the galaxy scale, which is physically unreasonable. The mass-to-light ratio of disk for ESO383G02 in the core-modified model is unphysically small. The mass-to-light ratios of ESO323G25, ESO383G02, ESO446G01 and ESO569G17 are fixed to be zero when fitting to the NFW model. MOND1 and MOND2 stand for the MOND models with standard and simple interpolation functions, respectively.}}
%\arrayrulewidth=1.0pt
\renewcommand{\arraystretch}{1.8}
\resizebox{!}{6.0cm}
 {\begin{tabular}{lccccccccccccccccccc} % creating cc10 columns
\hline\hline % inserting double-line
& \multicolumn{5}{c}{NFW} & \multicolumn{5}{c}{Core} \\
\cline{2-11}
 & $\sigma_{{\rm NFW}}$ & $\tau_{{\rm NFW}}$ & $M_{\rm vir}$ & $R_{\rm vir}$ &  $\chi^{2}_{{\rm NFW}}/{\rm dof}$ & $\sigma_{{\rm core}}$ & $\tau_{{\rm core}}$ & $\rho_{c}$ & $r_{c}$ &  $\chi^{2}_{{\rm core}}/{\rm dof}$\\
                &$M_{\odot}/L_{\odot}$ &$M_{\odot}/L_{\odot}$ & $10^{11}M_{\odot}$ &kpc &  &$M_{\odot}/L_{\odot}$ &$M_{\odot}/L_{\odot}$ & $M_{\odot}/{\rm pc}^{3}$ &kpc &  \\\hline
ESO215G39&	--&	--&	--&	--&	--&	--&	--&	--&	--&	--\\
ESO322G76&	--&	--&	--&	--&	--&	1.25$\pm$0.12&	1.35$\pm$0.10&	$(0.79\pm0.23)\times10^{-2}$&	13.45$\pm$4.06&	0.23\\
ESO322G77&-- &-- &-- &-- &-- &--&	--&	--&	--&	--\\
ESO323G25&	-- &	0.27$\pm$0.11&	6.14$\pm$0.08&	173.45$\pm$8.88&	0.16&  --&	--&	--& --&	--\\
ESO383G02&	--&	0.92$\pm$0.63&	2.95$\pm$0.16&	135.80$\pm$30.76&	0.14&--&	--&	--&	--&	--\\
ESO445G19&	2.13$\pm$1.64&	0.38$\pm$0.75&	11.81$\pm$1.68&	215.66$\pm$128.57&	0.08&4.29$\pm$0.67&	1.96$\pm$0.13&	$(0.54\pm0.23)\times10^{-2}$&	17.87$\pm$14.34&	0.08\\
ESO446G01&	--&	2.39$\pm$0.28&	0.68$\pm$0.06&	83.09$\pm$33.34&	0.51&0.94$\pm$0.34&	2.73$\pm$0.13&	20.99$\pm$9.12&	0.23$\pm$0.05&	0.52\\
ESO509G80&	1.25$\pm$0.60&	0.71$\pm$0.47&	30.85$\pm$1.98&	297.01$\pm$79.68&	0.20& --&	--&	-- &	--&	--\\
ESO569G17&	--&	1.40$\pm$0.38&	1.63$\pm$0.25&	111.51$\pm$72.30&	0.13&--&	--&	--&	--&	--\\
\hline
&&&&&&&&&&\\
\hline\hline
& \multicolumn{3}{c}{Baryon}  & \multicolumn{3}{c}{MOND1} & \multicolumn{3}{c}{MOND2} & \multicolumn{3}{c}{MSTG}\\
\cline{2-13}
    &$\sigma_{N}$ & $\tau_{N}$ & $\chi^{2}_{N}/{\rm dof}$ & $\sigma_{\rm MOND1}$ & $\tau_{\rm MOND}1$  & $\chi^{2}_{\rm MOND1}/{\rm dof}$ &
     $\sigma_{\rm MOND2}$ & $\tau_{\rm MOND2}$  & $\chi^{2}_{\rm MOND2}/{\rm dof}$ &$\sigma_{\rm MSTG}$ & $\tau_{\rm MSTG}$  & $\chi^{2}_{\rm MSTG}/{\rm dof}$\\
          &$M_{\odot}/L_{\odot}$ & $M_{\odot}/L_{\odot}$ & & $M_{\odot}/L_{\odot}$ & $M_{\odot}/L_{\odot}$  &  & $M_{\odot}/L_{\odot}$ & $M_{\odot}/L_{\odot}$ & & $M_{\odot}/L_{\odot}$ & $M_{\odot}/L_{\odot}$ &  &\\\hline
ESO215G39&	0.43$\pm$0.24&	1.81$\pm$0.05&	0.36&	0.79$\pm$0.18&	1.21$\pm$0.04&	0.16&	0.58$\pm$0.14&	0.86$\pm$0.03&	0.17
&1.26$\pm$0.19&	1.24$\pm$0.03&	0.20\\
ESO322G76&	0.75$\pm$0.21&	1.92$\pm$0.08&	1.19&	1.15$\pm$0.11&	1.31$\pm$0.04&	0.29&	0.97$\pm$0.10&	0.93$\pm$0.03&	0.31
&1.36$\pm$0.10 &1.19$\pm$0.03 &0.25 \\
ESO322G77 &1.47$\pm$0.51 &3.20$\pm$0.11	&0.33	&1.71$\pm$0.45	& 2.98$\pm$0.10	&0.24	&1.77$\pm$0.40	&2.31$\pm$0.08	&0.20
&2.23$\pm$0.41	&2.75$\pm$0.08	&0.20 \\
ESO323G25&	5.15$\pm$1.92&	2.67$\pm$0.07&	1.51&7.96$\pm$0.85&	2.14$\pm$0.03&	0.26&	8.31$\pm$0.96&	1.57$\pm$0.03&	0.34
&11.14$\pm$1.38&	1.70$\pm$0.04&	0.65\\
ESO383G02&	3.23$\pm$0.33&	3.05$\pm$0.07&	0.17&5.56$\pm$0.52&	1.70$\pm$0.11&	0.36&	4.55$\pm$0.39&	1.13$\pm$0.08&	0.31
&6.69$\pm$0.48&	1.26$\pm$0.10&	0.50\\
ESO445G19&	1.66$\pm$0.83&	2.48$\pm$0.06&	0.27&3.85$\pm$0.50&	1.81$\pm$0.03&	0.08&	3.22$\pm$0.41&	1.28$\pm$0.03&	0.08
&6.48$\pm$0.57&	1.55$\pm$0.03&	0.13\\
ESO446G01&	2.78$\pm$0.25&	2.34$\pm$0.22&	1.65&3.20$\pm$0.20&	1.09$\pm$0.18&	1.02&	2.68$\pm$0.19&	0.67$\pm$0.14&	1.14
&3.16$\pm$0.18&	0.80$\pm$0.14&	1.07\\
ESO509G80&	0.40$\pm$0.43&	3.21$\pm$0.09&	0.82&1.38$\pm$0.28&	2.54$\pm$0.05&	0.28&	1.35$\pm$0.23&	1.82$\pm$0.04&	0.24
&2.98$\pm$0.24&	1.93$\pm$0.04&	0.26\\
ESO569G17&	0.12$\pm$0.21&	2.10$\pm$0.06&	0.19&0.16$\pm$0.16&	1.87$\pm$0.05&	0.11&	0.34$\pm$0.16&	1.40$\pm$0.05&	0.13
&0.50$\pm$0.16&	1.71$\pm$0.05&	0.13\\
\hline\hline% inserts single-line
\end{tabular}}\label{tab:HSB_parameter}
\end{table*}

\begin{table*}[htbp]
\centering                  %表格居中
\caption{\small{The best-fitting parameters of LSB galaxies in different models. The parameters of the NFW model couldn't be well constrained. In the core-modified profile, the errors of mass-to-light ratios of disk for three galaxy (F561-1, F583-1, UGC5750) overstep central values. The mass-to-light ratios of disk for F568-3 in the core-modified model, and for F561-1 in MOND2 model are unphysically small. MOND1 and MOND2 stand for the MOND models with standard and simple interpolation functions, respectively.}}
%\arrayrulewidth=1.0pt
\renewcommand{\arraystretch}{1.5}
\resizebox{!}{2.5cm}
 {\begin{tabular}{lccccccccccccccc} % creating cc10 columns
\hline\hline % inserting double-line
& \multicolumn{2}{c}{Baryon}   & \multicolumn{2}{c}{MOND1} & \multicolumn{2}{c}{MOND2} & \multicolumn{2}{c}{MSTG} &\multicolumn{4}{c}{Core}\\
\cline{2-13}
    & $\tau_{N}$ & $\chi^{2}_{N}/{\rm dof}$ &  $\tau_{\rm MOND1}$ & $\chi^{2}_{\rm MOND1}/{\rm dof}$ &  $\tau_{\rm MOND2}$ & $\chi^{2}_{\rm MOND2}/{\rm dof}$&  $\tau_{\rm MSTG}$ & $\chi^{2}_{\rm MSTG}/{\rm dof}$ &$\tau_{core}$ &$\rho_{c}$ &$r_{c}$ &$\chi^{2}_{\rm Core}/{\rm dof}$\\
                      & $M_{\odot}/L_{\odot}$ &   & $M_{\odot}/L_{\odot}$  &  & $M_{\odot}/L_{\odot}$  &  & $M_{\odot}/L_{\odot}$  &  &$M_{\odot}/L_{\odot}$ & $M_{\odot}/{\rm pc}^{3}$ &kpc & \\\hline
F561-1&	2.56$\pm$0.23&	0.66&		0.03$\pm$0.20&	11.24&-- &-- &	1.06$\pm$1.23&	26.22 &--&	--&	--&--\\
F563-1&	23.00$\pm$3.26&	3.00&		4.46$\pm$0.26&	0.07& 3.26$\pm$0.19&	0.07
	&4.97$\pm$0.87&	0.65 &9.50$\pm$1.46&	$(0.41\pm0.13)\times10^{-2}$&	10.94$\pm$1229&	0.08\\
F568-3&	6.09$\pm$0.64&	2.15&		1.62$\pm$0.32&	1.63&	1.24$\pm$0.24&1.71
&3.49$\pm$0.51& 2.32& --&	--&	--&	--\\
F579-V1&8.55$\pm$0.63&	1.40&	3.71$\pm$0.69&	2.58&	2.72$\pm$0.48&	2.57
&5.04$\pm$0.94&	5.48 &5.79$\pm$0.50&	0.40$\pm$0.11&	0.77$\pm$0.13&	0.16\\
F583-1&	7.12$\pm$1.51&	8.43& 1.50$\pm$0.43&	3.36&	1.19$\pm$0.37&	4.35
&6.03$\pm$1.79&	15.46& --&	--&	--&	--\\
F730-V1&   7.71$\pm$0.53&	1.99&		4.18$\pm$0.55&	2.53&	3.02$\pm$0.37&	2.46
&5.34$\pm$0.87&	8.12&5.82$\pm$0.41&	0.11$\pm$0.02&	1.39$\pm$0.21&	0.15\\
UGC128&	6.30$\pm$1.34&	8.73&	1.00$\pm$0.14&	0.57&	0.79$\pm$0.10&	0.55
&2.86$\pm$0.54&	2.87&3.19$\pm$0.23&	$(0.34\pm0.05)\times10^{-2}$&	13.97$\pm$1.06&	0.09\\
UGC1230&	4.07$\pm$1.24&	11.97&		0.18$\pm$0.11&	1.83&	0.14$\pm$0.10&	2.26
&2.67$\pm$1.10&	13.94&1.45$\pm$0.30&	$(0.49\pm0.09)\times10^{-2}$&	9.01$\pm$0.90&	0.26\\
UGC5750&	1.58$\pm$0.21&	1.90&	0.07$\pm$0.03&	0.57&	0.05$\pm$0.02&	0.55
&0.62$\pm$0.13&	1.61&--&	--&	--&	--\\\hline
\hline% inserts single-line
\end{tabular}}\label{tab:LSB_parameter}
\end{table*}

%\ruleup

\section{Model comparison}\label{sec:comparison}

\begin{figure*}[htbp]
\centering
\includegraphics[width=0.5\textwidth]{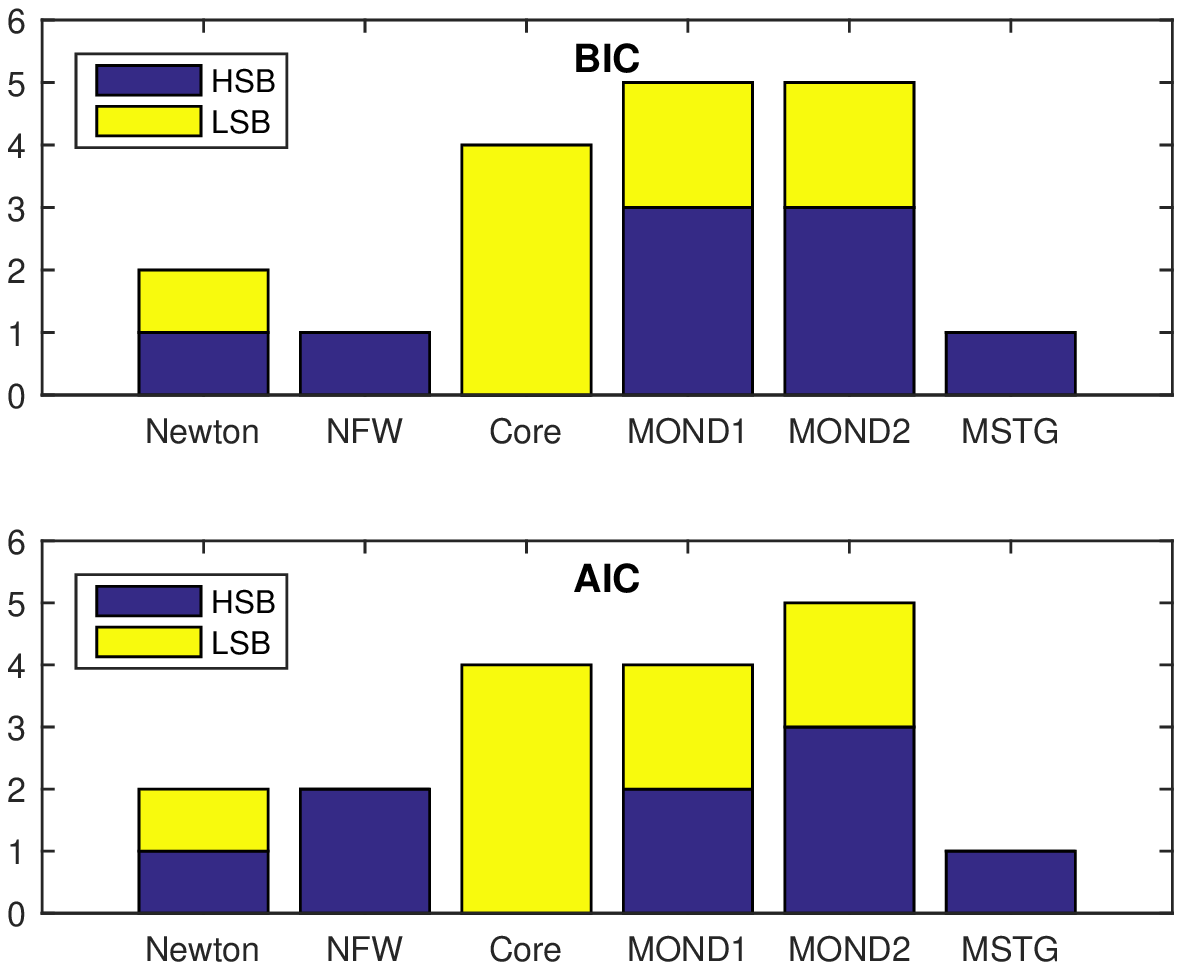}
\caption{The number of galaxies that can be best fitted by each model. In the upper panel BIC is applied, while in the lower panel AIC is applied.}\label{fig:barplot}
\end{figure*}

To appraise which model is the best, one may adopt the most direct method by comparing the $\chi^{2}$ of each model, thereby reveals that the model whose $\chi^{2}$ is the smallest is the best. However, a model with more parameters in general has smaller $\chi^2$. Because the dark matter models have two more free parameters than other models, the result is not comprehensive. One may prefer to use the reduced-$\chi^2$, i.e. the $\chi^2$ per degree of freedom to measure the goodness of fit.  We presented the reduced-$\chi^{2}$ in Table~\ref{tab:HSB_parameter} and Table~\ref{tab:LSB_parameter} for HSB and LSB galaxies, respectively. However, the reduced-$\chi^2$ is still not comprehensive enough to depict models. Therefore, in this section, we compare models with statistical analysis based on the likelihood function $\mathcal{L}=\exp\left(-\chi^{2}/2\right)$. One of the most used criteria to describe the goodness-of-fit is the Bayesian Information Criterion (BIC) \cite{Schwarz:1978},
\begin{equation}
{\rm BIC}=-2{\rm ln}\mathcal{L}_{\rm max}+p{\rm ln}N.
\end{equation}
where $N$ is the number of data points in the galaxy rotation curve, and $p$ is the number of free parameters. Another widely used criterion is the Akaike Information Criterion (AIC) \cite{Akaike:1974},
\begin{equation}
{\rm AIC}=-2{\rm ln}\mathcal{L}_{\rm max}+2p.
\end{equation}
We list the $\chi^2$, BIC and AIC values for each model in Table~\ref{tab:comparison}. Models with the smallest value of BIC or AIC highlighted in boldface are the best models. To be more visible, in Fig.~\ref{fig:barplot} we plot the number of galaxies that can be best fitted by each model.

\begin{table*}[htbp]
\centering                  %表格居中
\caption{\small{The statistical comparison between different models. The best models are highlighted in boldface.}}
%\arrayrulewidth=1.0pt
\renewcommand{\arraystretch}{1.5}
\resizebox{!}{3.5cm}
 {\begin{tabular}{lcccccccccccccccccc} % creating cc10 columns
\hline\hline % inserting double-line
  & $\chi^{2}_{N}$  & $\chi^{2}_{\rm NFW}$ & $\chi^{2}_{\rm Core}$ & $\chi^{2}_{\rm MOND1}$& $\chi^{2}_{\rm MOND2}$ & $\chi^{2}_{\rm MSTG}$ & ${\rm BIC}_{N}$ &${\rm BIC}_{\rm NFW}$ &${\rm BIC}_{\rm Core}$ &${\rm BIC}_{\rm MOND1}$ &${\rm BIC}_{\rm MOND2}$& ${\rm BIC}_{\rm MSTG}$ &${\rm AIC}_{N}$ &${\rm AIC}_{\rm NFW}$&${\rm AIC}_{\rm Core}$&${\rm AIC}_{\rm MOND1}$&${\rm AIC}_{\rm MOND2}$&${\rm AIC}_{\rm MSTG}$ \\\hline
ESO215G39&	12.31& 	--& 	--& 	5.38& 5.61&	6.96& 	19.48& 	--& 	--& 	\bf{12.54}& 12.78&	14.12& 	16.31& 	--& 	--& 	\bf{9.38}&9.61& 	10.96\\
ESO322G76&	62.97& 	--& 	11.8&	15.15& 16.33 &	13.33& 	70.98& 	--& 	27.8& 	23.16& 24.34&	\bf{21.35}& 	66.97& 	--& 	19.8& 	19.15& 20.33&	\bf{17.33}\\
ESO322G77&	4.58& 	--& 	--& 	3.38& 2.83&	2.85& 	10.13& 	--& 	--& 	8.92& \bf{8.38}&	8.40& 	8.58& 	--& 	--& 	7.38& \bf{6.83}&	6.85\\
ESO323G25&	97.85& 	10.38& 	--& 	16.77& 22.30&	42.16& 	106.26&	27.19& 	--& 	\bf{25.18}& 30.71&	50.57& 	101.85&	\bf{18.38}& 	--& 	20.77&26.30& 	46.16\\
ESO383G02&	6.76& 	5.48& 	--& 	14.32& 12.22& 	20.18& 	\bf{14.24}& 	20.43& 	--& 	21.79& 19.70&	27.65& 	\bf{10.76}& 	13.48& 	--& 	18.32& 16.22&	24.18\\
ESO445G19&	10.24& 	2.75& 	2.87& 	3.00& 2.98&	4.76& 	17.62& 	17.51& 	17.62& 	10.38& \bf{10.36}&	12.14& 	14.24& 	10.75& 	10.87& 	7.00& \bf{6.98}&	8.76\\
ESO446G01&	72.77& 	21.45& 	22.00& 	44.91& 50.27&	47.13& 	80.42& 	\bf{36.77}& 	37.31& 	52.56& 57.92&	54.79& 	76.77& 	\bf{29.45}& 	30.00& 	48.91& 54.27&	51.13\\
ESO509G80&	29.51& 	6.70& 	--& 	10.06& 8.74&	9.42& 	36.79& 	21.26& 	--& 	17.33& \bf{16.02}&	16.69& 	33.51& 	14.70& 	--& 	14.06& \bf{12.74}&	13.42\\
ESO569G17&	3.81& 	2.27& 	--& 	2.29& 2.51&	2.64& 	9.99& 	14.64& 	--& 	\bf{8.47}& 8.70 &	8.82& 	7.81& 	10.27& 	--& 	\bf{6.29}& 6.51&	6.64\\
F561-1&	    3.97& 	--& 	--& 	67.41& --&	157.31&	\bf{5.92}& 	--& 	--& 	69.35& --&	159.25&	\bf{5.97}& 	--& 	--& 	69.41& --&  159.31\\
F563-1&	    27.04& 	--& 	0.58& 	0.67& 0.63&	5.85& 	29.34& 	--& 	7.49& 	2.97& \bf{2.94}&	8.15& 	29.04& 	--& 	6.58& 	2.67& \bf{2.63}&	7.85\\
F568-3&	    21.51& 	--& 	--& 	16.27& 17.08&	23.19& 	23.91& 	--& 	--& 	\bf{18.66}& 19.47&	25.58& 	23.51& 	--& 	--& 	\bf{18.27}&19.08& 	25.19\\
F579-V1&    18.25& 	--& 	1.78& 	33.58& 33.41&	71.25& 	20.89& 	--& 	\bf{9.70}& 	36.22&36.05& 	73.89& 	20.25& 	--& 	\bf{7.78}& 	35.58& 35.41&	73.25\\
F583-1&	    134.92&	--& 	--&	53.72& 69.55&	247.40&	137.76&	--& 	--& 	\bf{56.56}& 72.39&	250.24&	136.92&	--& 	--& 	\bf{55.72}& 71.55& 249.40\\
F730-V1&    13.92& 	--& 	0.76& 	17.72& 17.19&	56.84& 	16.00& 	--& 	\bf{7.00}& 	19.80& 19.27&	58.92& 	15.92& 	--& 	\bf{6.76}& 	19.72& 19.19&	58.84\\
UGC128&	    96.06& 	--& 	0.85& 	6.22& 6.05&	31.57& 	98.54& 	--& 	\bf{8.31}& 	8.70& 8.54&	34.05& 	98.06& 	--& 	\bf{6.85}& 	8.22& 8.05&	33.57\\
UGC1230&    119.73&	--& 	2.09& 	18.31& 22.64&	139.45&	122.13&	--& 	\bf{9.28}& 	20.71& 25.04&	141.85&	121.73&	--& 	\bf{8.09}& 	20.31& 24.64&141.45\\
UGC5750&    18.96& 	--& 	--& 	5.75& 5.53&	16.06& 	21.36& 	--& 	--& 	8.15& \bf{7.93}&	18.45& 	20.96& 	--& --& 7.75& \bf{7.53}&  18.06\\\hline
\hline% inserts single-line
\end{tabular}}\label{tab:comparison}
\end{table*}

According to the BIC criterion, one HSB galaxy (ESO383G02) and one LSB galaxie (F561-1) are best fitted by the Baryonic model. Only one HSB galaxy (ESO446G01) but no LSB galaxy is best fitted by the NFW model. One HSB galaxy (ESO322G76) but no LSB galaxy is best fitted by the MSTG model. Four LSB galaxies (F579-V1, F730-V1, UGC128 and UGC1230) but no HSB galaxy are best fitted by the core-modified profile. Simple MOND model fits well for three HSB galaxies (ESO322G77, ESO445G19 and ESO509G80) and two LSB galaxies (F563-1 and UGC5750). For the rest three HSB and two LSB galaxies, standard MOND model is the best model.

If we apply the AIC criterion, similar conclusions can be arrived. The only difference between BIC and AIC happens in the HSB galaxy ESO323G25, which according to the BIC criterion is best fitted by standard MOND model, while according to the AIC criterion is best fitted by NFW model. In fact, ESO323G25 can be fitted by both models very well.

\section{Discussion and summary}\label{sec:summary}

In this paper, we have compared six different models (Baryonic model, NFW profile, core-modified profile, standard MOND, simple MOND and MSTG) in account for the rotation curves of 9 HSB and 9 LSB galaxies. We fitted the observed rotation curve data to theoretical models, and used the Bayesian Information Criterion (BIC) and Akaike Information Criterion (AIC) to appraise which model is the best. We found that non of the six models can well fit all the 18 galaxies. Specifically, non of the HSB galaxies can be well fitted by core-modified model, and non of the LSB galaxies can be well fitted by NFW model. Only one or two (depends on either BIC or AIC is applied) HSB galaxies are best accounted by NFW model. This hits that the dark matter halos, if they really exist, in some cases couldn't be well mimicked by the oversimplified NFW or core-modified profiles. Among the 18 galaxies, only one HSB galaxy can be best fitted by MSTG model, which implies that MSTG is not a universe model. Two galaxies (one HSB galaxy and one LSB galaxy) are best accounted by Baryonic model. For these two galaxies, it is neither necessary to add the dark matter component, nor necessary to modified the Newtonian dynamics or Newtonian gravity. Three or two HSB galaxies are best met by standard MOND model, and three HSB galaxies are best fitted by simple MOND model. In most case, standard MOND and simple MOND fit the data equally well. In summary, we couldn't arrive a convincing conclusion to prefer one model and exclude the others.

\acknowledgments{We are grateful to J. Li, H. Ma and L. L. Wang for useful discussions. This work has been supported by the Fundamental Research Funds for the Central Universities (Grant No. 106112016CDJCR301206), the National Natural Science Fund of China (Grant Nos. 11305181 and 11547305), and the Open Project Program of State Key Laboratory of Theoretical Physics, Institute of Theoretical Physics, Chinese Academy of Sciences, China (Grant No. Y5KF181CJ1).}

\end{document}